\documentclass[final,5p,times,twocolumn,authoryear]{elsarticle}

\makeatletter
\def\ps@pprintTitle{%
  \let\@oddhead\@empty
  \let\@evenhead\@empty
  \def\@oddfoot{\reset@font\hfil\thepage\hfil}
  \let\@evenfoot\@oddfoot
}
\makeatother

\usepackage{graphicx}%
\usepackage{multirow}%
\usepackage{amsmath,amssymb,amsfonts}%
\usepackage{amsthm}%
\usepackage{mathrsfs}%
\usepackage[title]{appendix}%
\usepackage{xcolor}%
\usepackage{textcomp}%
\usepackage{manyfoot}%
\usepackage{booktabs}%
\usepackage{algorithm}%
\usepackage{algorithmicx}%
\usepackage{algpseudocode}%
\usepackage{listings}%
\usepackage{tabularx}
\usepackage{fancyhdr}
\usepackage{datetime}

\usepackage{fancyhdr}
\journal{SSRN}

\begin{document}

\begin{frontmatter}



\title{Understanding gender differences in experiences and concerns surrounding online harms: A short report on a nationally representative survey of UK adults}

\affiliation[label1]{organization={Public Policy Programme, The Alan Turing Institute},
city={London},
country={UK}}

\affiliation[label2]{organization={Oxford Internet Institute, University of Oxford},
city={Oxford},
country={UK}}

\author[label1]{Florence E. Enock}
\author[label1]{Francesca Stevens}
\author[label1]{Jonathan Bright}
\author[label1]{Miranda Cross}
\author[label1]{Pica Johansson}
\author[label1]{Judy Wajcman}
\author[label1,label2]{Helen Z. Margetts}

\begin{abstract}
Online harms, such as hate speech, misinformation, harassment and self-harm promotion, continue to be widespread. While some work suggests that women are disproportionately affected by such harms, other studies find little evidence for gender differences in overall exposure. Here, we present preliminary results from a large, nationally representative survey of UK adults (N = 2000). We asked about exposure to 15 specific harms, along with fears surrounding exposure and comfort engaging in certain online behaviours. While men and women report seeing online harms to a roughly equal extent overall, we find that women are significantly more fearful of experiencing every type of harm that we asked about, and are significantly less comfortable partaking in several online behaviours. Strikingly, just 24\% of women report being comfortable expressing political opinions online compared with almost 40\% of men (translating to women being 50\% less comfortable than men participating politically online), with similar overall proportions for challenging certain content. Our work suggests that women may suffer an additional psychological burden in response to the proliferation of harmful online content, doing more 'safety work' to protect themselves. With much public discourse happening online, gender inequality in public voice is likely to be perpetuated if women feel too fearful to participate. Our results are important because to establish greater equality in society, we must take measures to ensure all members feel safe and able to participate in the online space. 
\end{abstract}




\end{frontmatter}




\section{Background}
\label{background}
While social media has the potential to bring benefits to its users, such as fostering social connection, helping maintain long distance friendships, facilitating education and empowering individuals with a platform for their voice, it is also without doubt that social media and the online world more broadly has the potential to expose people to many forms of harm. Harms people may be exposed to online include hate speech, abuse, bullying, misinformation, stalking, image based sexual abuse and self-harm promotion, amongst many others. Recent work has found that 66\% of adults in Britain have seen content which they consider to be harmful online before, while for younger adults aged 18-34, this is 86\% \citep{Turing2023_harmstracker}. Some work suggests that women and girls are disproportionately affected by experiences of online harms: for example, a report from the European Women's Lobby found that the internet was a place of gendered violence \citep{EWL2017}. However, there is conflicting evidence regarding gender differences in absolute exposure to online harms, with some recent work finding no differences in overall exposure \citep{Turing2023_harmstracker, OfCom2023}. 

To deepen our understanding of gender differences in how people are affected by online harms, we conducted a nationally representative survey of UK adults, asking about experiences and concerns surrounding 15 specific types of online harm. In this short report, we present key findings showing gender differences in experiences of each harm, fears about being exposed to each harm, and comfort with participating in certain online behaviours such as expressing political opinions, sharing personal information and challenging certain content.  

\section{Data and Methods}
\label{Data and Methods}
We conducted a nationally representative survey of 2000 UK adults. The survey was created and administered using Qualtrics\footnote{www.qualtrics.com} and participants were recruited through Prolific\footnote{https://www.prolific.com}. Data was collected during June 2023 and total of 1,992 participants who completed the survey passed standard checks for data quality and were included in the final analysis. The sample was designed to be nationally representative of the population of the United Kingdom across demographic variables of age, gender and ethnicity. Respondents were between 18 and 90 years old, with a mean age of 45.7 (SD = 15.5). A total of 1010 participants identified as female (50.7\%), 965 as male (48.4\%), with 9 as non-binary, 5 selecting ‘prefer not to say’, and 3 self-describing. 

The survey was approved by the Ethics Committee at The Alan Turing Institute, UK (approval number C2105-074). Informed consent was obtained at the start of the survey according to approved ethical procedures. Here, we report responses to three key questions that were included in the survey. 

\textbf{Exposure to online harms}:
We first presented participants with a list of 15 online harms along with definitions for each. The harms we included were: Hate speech, Misinformation, Misogyny, Trolling, Bullying, Cyberstalking, Cyberflashing, Group attacks, Impersonation, Catfishing, Threats of non-sexual violence, Threats of sexual violence, Doxing, Image based sexual abuse, and Eating disorder content. Participants indicated whether they had heard of or seen each harm online in the past year (Never heard of/ Heard of but not seen in the past year/Seen online in the past year). 

\textbf{Fears of exposure to online harms}:
To measure concerns about online harms, for each of the same 15 harms, participants indicated the extent to which they fear witnessing such content (for example, in posts not directly targeted at them), and the extent to which they fear directly receiving such content (for example, in posts or messages directly intended for them) (both scales: Not at all / Not very much / Somewhat / Very much). 

\textbf{Comfort with online activities}:
We asked participants how comfortable they are (Not at all/Not very/Somewhat/Extremely) with engaging in a variety of online behaviours in both public and private settings. Here, we report responses for publicly expressing political opinions, expressing other opinions (for example, about films or music), sharing personal information, and challenging content that the respondent does not agree with. 

We present descriptive statistics for overall proportions of men and women choosing each response option for each question \footnote{While 12 participants selected a gender identity other than male or female, and 5 chose not to respond to the gender identity question, these numbers are too small to make meaningful gender comparisons and for the purposes of the current research we only report responses from respondents identifying as male or female.}. For each of the three outcome variables of interest, we use logistic regressions to test for gender differences with Gender (Male/Female) as predictor and outcomes as: 1. Seen/Not seen for each of the 15 harms; 2. Fearful/Not fearful for each of the 15 harms; and 3. Comfortable/Not comfortable with participating in each of the 4 online behaviours. The accepted significance threshold was set at .05. 

\section{Results}
\label{Results}
\subsection{Exposure to online harms}
We present responses for awareness of and exposure to the 15 online harms we included in the survey questions for men and women. Here, we count exposure by a response indicating `Seen online in the last year'. Gender differences in exposure to various harms show mixed results. Men report significantly greater exposure to some harms such as hate speech, misinformation, impersonation, threats of violence and doxing, while women report significantly greater exposure to other forms of harm such as online misogyny, bullying, and eating disorder content. For example, 64\% of men say they have seen online hate speech compared to 59\% of women and 36\% of men say they have seen threats of physical violence compared to 28\% of women, while 54\% of women say they have seen misogyny online compared to 50\% of men, and 28\% of women say they have seen eating disorder content online compared to 16\% of men. This translates as men being 18\% more likely to see hate speech and 31\% more likely to see threats of physical violence than women, while women are 21\% more likely to see online misogyny and 109\% more likely to see eating disorder content than men. For many of the harms, such as trolling and cyberstalking, differences in self-reported exposure between men and women are not statistically significant. Figure 1 shows self-reported exposure to the 15 harms by gender.

\subsection{Fears about exposure to online harms}
Comparing gender differences in fears about exposure to each of the 15 online harms, we created a binary outcome for fear, with `Somewhat' and `Very much' as Fearful, and `Not at all' and `Not very much' as Not fearful. Women consistently express significantly greater levels of fear than men across all 15 types of content, both for witnessing and for directly receiving. For example, 49\% of women fear receiving misogynistic content compared to 14\% of men, 54\% of women fear being the target of cyberstalking compared to 35\% of men, 45\% of women fear being targeted by cyberflashing compared to 21\% of men, and 50\% of women fear being the target of image based sexual abuse compared to 30\% of men. These results translate as women being 505\% more fearful of receiving misogynistic content, 119\% more fearful of being the target of cyberstalking, 197\% more fearful of being targeted by cyberflashing, and 136\% more fearful of being targeted by image based sexual abuse than men. Figure 2 shows self-reported fear about receiving each of the 15 harms by gender, with all gender differences statistically significant at the .05 level. 

\subsection{Comfort with online behaviours}
To test for gender differences in comfort with four public online behaviours (expressing political opinions, expressing opinions more generally, sharing personal information and challenging disagreeable content), we created a binary outcome for comfort, with `Somewhat' and `Extremely' as Comfortable, and `Not at all' and `Not very' as Not comfortable. Women express significantly lower levels of comfort with engaging in these behaviours than men across all four examples. Only 24\% of women are comfortable expressing political opinions online compared with almost 40\% of men, 51\% of women are comfortable expressing other opinions (for example, about the news, movies or music) online compared with 65\% of men, 8\% of women are comfortable sharing personal information online compared with 11\% of men, and just 23\% of women are comfortable with challenging content they do not agree with online, compared with 40\% of men. This means that women are 51\% less comfortable than men expressing political opinions, 45\% less comfortable than men expressing other opinions, 34\% less comfortable than men sharing personal information, and 55\% less comfortable than men challenging content they disagree with. Figure 3 shows self-reported comfort with four online behaviours, with all gender differences significant. 

 \begin{figure*}
    \centering
    \includegraphics[width=1\linewidth]{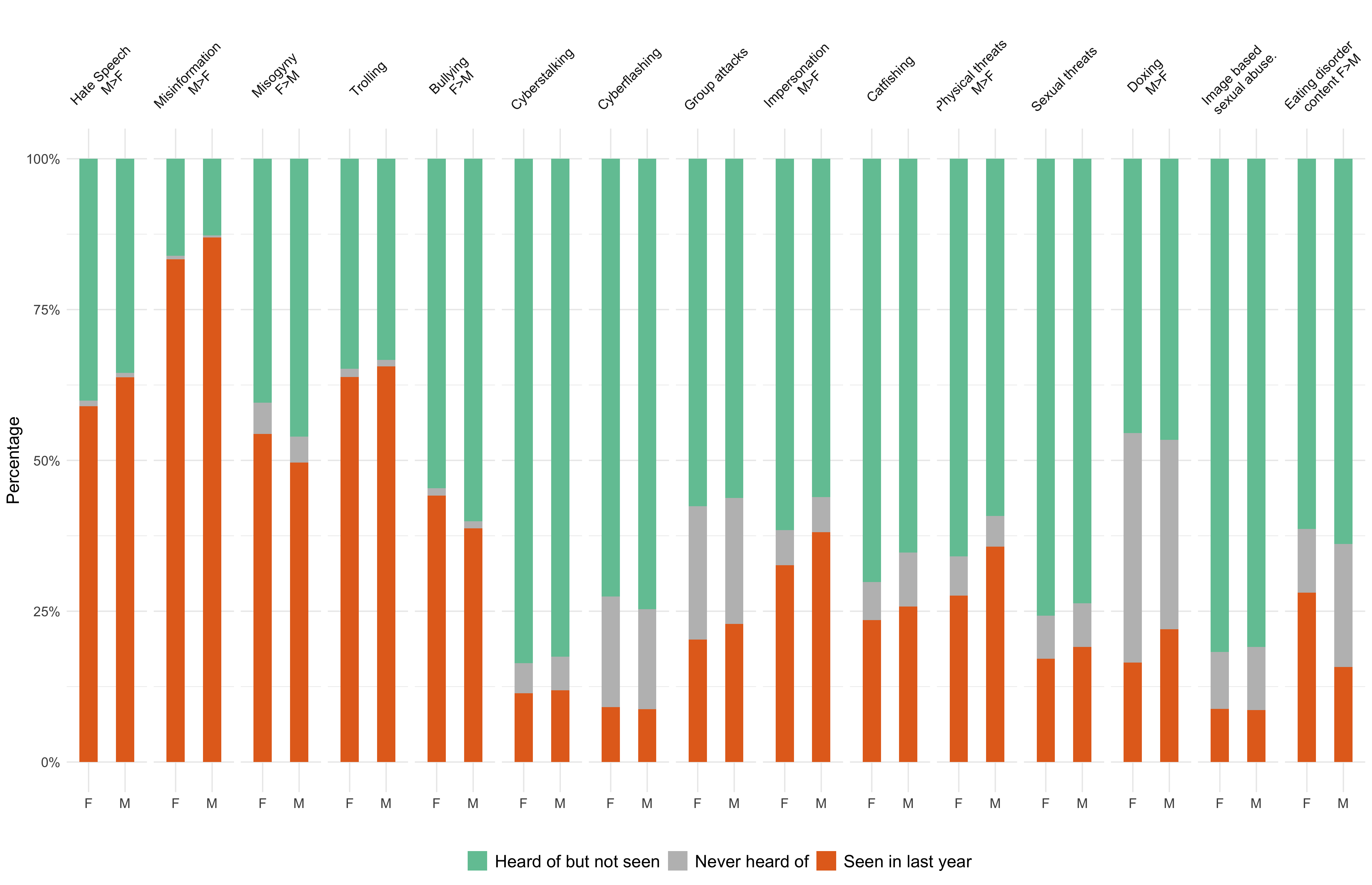}
    \caption{Self-reported exposure to 15 online harms by gender. F$>$M indicates self-reported exposure is significantly higher for women, M$>$F indicates self-reported exposure is significantly higher for men}
    \label{fig:exposure}
\end{figure*}

\begin{figure*}
    \centering
    \includegraphics[width=1\linewidth]{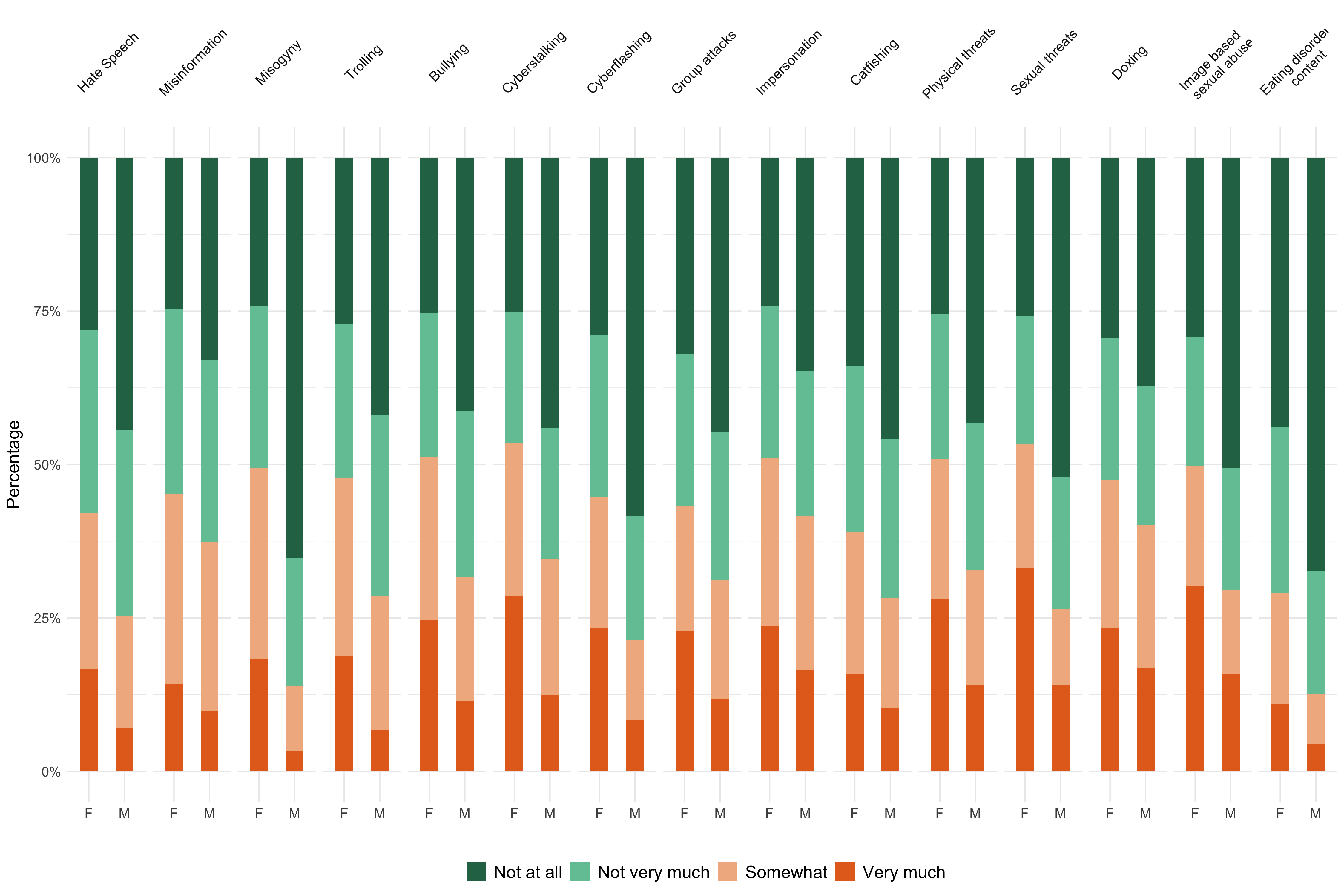}
    \caption{Self-reported fear of receiving harmful content online by gender. All gender differences are statistically significant at the .05 level}
    \label{fig:fear-recieving}
\end{figure*}

\begin{figure*}
    \centering
    \includegraphics[width=1\linewidth]{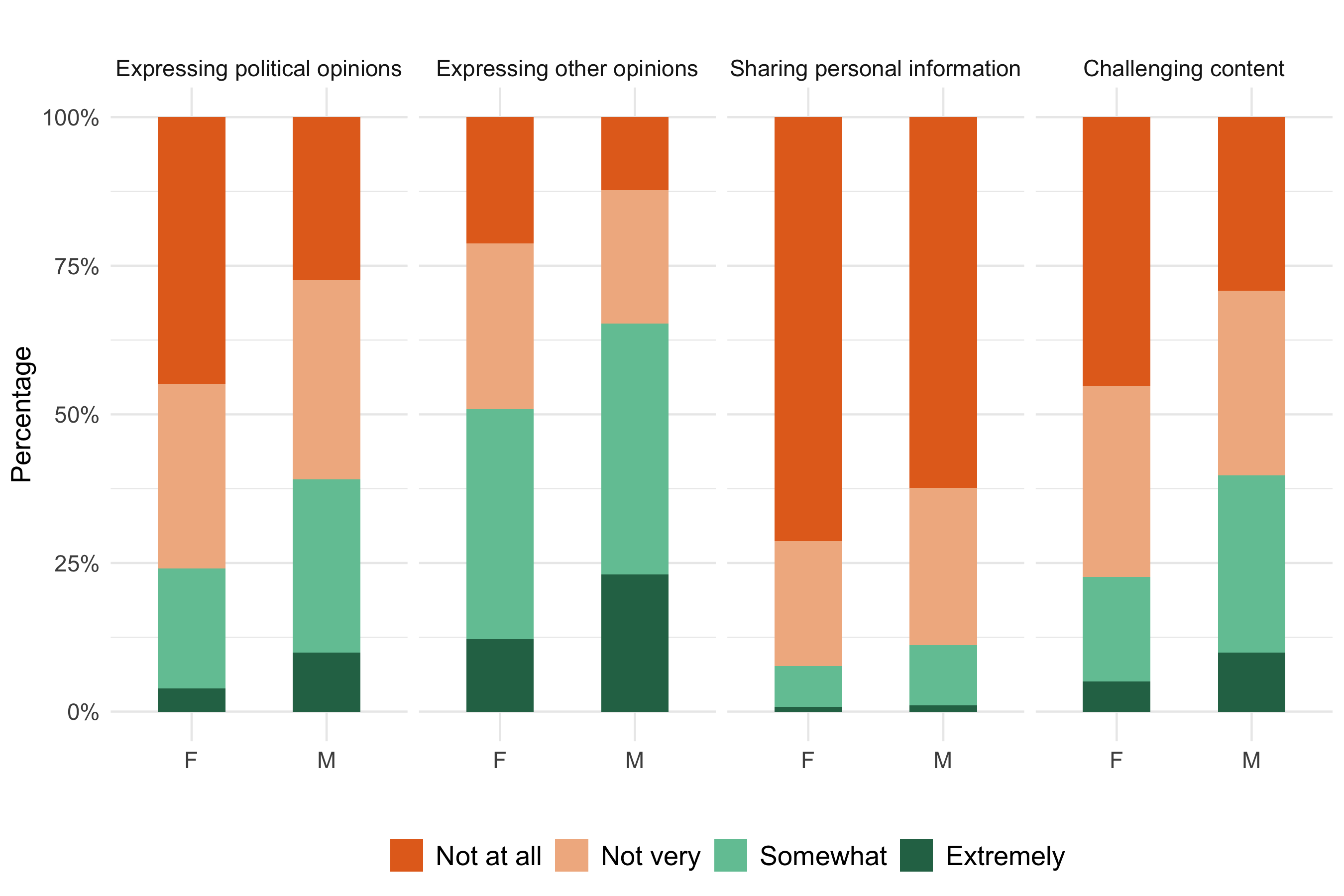}
    \caption{Comfort with four online behaviours by gender. All gender differences are significant at the .05 level}
    \label{fig:comfort_level}
\end{figure*}

\section{Conclusion}
\label{Conclusion}
In a large, nationally representative survey of 2000 adults in the UK, we examined gender differences in exposure to 15 specific online harms, along with fears surrounding these harms and comfort with particular online behaviours. We found little difference between men and women's self-reported exposure to online harm overall, in line with some other recent work \citep{Turing2023_harmstracker, OfCom2023}. For some of the harms, like hate speech and misinformation, men report greater exposure, while for others, such as misogyny and bullying, women report greater exposure. The work suggests that men and women may be more likely to see different kinds of online harms, but that exposure does not differ greatly overall. 

For each of the 15 harms we asked about, women express significantly more fear about seeing them than men do. In line with some related work from Amnesty International, this suggests an additional psychological burden on women in the online space \citep{Amnesty2017, EWL2017}. Correspondingly, women are significantly less comfortable than men partaking in online behaviours, here including expressing political opinions, expressing other opinions, sharing personal information, and challenging undesirable content. Although we do not test this link formally, it is plausible that greater fear surrounding exposure to online harms leads women to do more to protect themselves online, including self-censorship on political and other opinions. These results are concerning, because with much public discourse happening online, gender inequality in public voice is likely to be perpetuated if women feel too fearful to participate. 

The results we present here are preliminary, and future work will seek to deepen our understanding about where people's fears about online harms typically arise from, common psychological impacts of online experiences, and whether there are gender differences in engaging with additional safety behaviours. Importantly, while our work suggests that greater fear amongst women leads to increased `safety work' (perhaps successful in reducing exposure to some types of online harms that individuals would otherwise have experienced), future work will seek to understand this mechanism in more depth. To work towards a more equal society, we must ensure all members of society feel safe and able to participate in the online space. 

\section*{Corresponding author:}
\label{Author contact}
Florence E. Enock, fenock@turing.ac.uk 


\section*{Acknowledgements}
This work was supported by the Ecosystem Leadership Award under the EPSRC Grant EPX03870X1 and The Alan Turing Institute. 




\bibliographystyle{apalike} 
\bibliography{Survey3_citations}





\end{document}